\documentclass[onecolumn,preprint,showpacs,showkeys,superscriptaddress]{revtex4-1}
\usepackage{amssymb}
\usepackage{amsfonts}
\usepackage{amsmath}
\usepackage{graphicx}
\usepackage{epstopdf}
\usepackage{float}

\begin{document}

\preprint{}
\title{Fermions in the Rindler spacetime}
\author{L. C. N. Santos}
\email{luis.santos@ufsc.br}
\affiliation{Departamento de F\'isica, CCEN - Universidade Federal da Paraíba; C.P. 5008, CEP  58.051-970, João Pessoa, PB, Brasil}
\author{C. C. Barros Jr.}
\email{barros.celso@ufsc.br}
\affiliation{Departamento de F\'{\i}sica - CFM - Universidade Federal de Santa Catarina, CP. 476
- CEP 88.040 - 900, Florian\'{o}polis - SC - Brazil}

\begin{abstract}
In this paper we study the Dirac equation in the Rindler spacetime. The
solution of the wave equation in an accelerated reference frame is obtained.
The differential equation associated to this wave equation is mapped into a Sturm-Liouville problem of a Schrödinger-like equation. We derive a compact expression for the energy spectrum associated with the Dirac
equation in an accelerated reference. It is shown that the
noninertial effect of the accelerated reference frame mimics an external
potential in the Dirac equation and, moreover, allows the formation of bound
states.
\end{abstract}
\keywords{Noninertial effects; relativistic bound states;  Dirac equation}

\pacs{03.65.Ge, 03.65.--w, 03.65.Pm, 04.20.Gz}

\eid{identifier}
\startpage{1}

\maketitle

\section{Introduction}

In recent years, the scientific interest in the study of noninertial effects
on physical systems has been renewed and many systems have been studied \cite%
{Castro1,Castro2,inercial9,inercial10,inercial7}. For instance, in \cite%
{Castro1} it was shown that the noninertial effect breaks the symmetrical of
the energy spectrum about $E=0$. In this way, it was shown that the energy
spectrum associated with scalar bosons in a rotating reference frame is
different from the one obtained in a usual inertial reference frame, in
other words, the energy levels are shifted by the effects of the rotating
frame.

Another kind of noninertial system that may be investigated with this
purpose are the uniformly accelerated observers in Minkowski spacetime, the
so-called Rindler spacetime. In order to investigate the energy states of
any quantized field in a accelerated reference frame it was suggested that a
uniformly accelerated detector in vacuum measures blackbody radiation, i.e,
an observer who undergoes a acceleration apparently sees a fixed surface
radiate \cite{uru1,uru2,uru3}.

Parallel to this, gravitational effects on quantum mechanical systems have
been studied intensively over the last few years \cite%
{parker1,parker2,parker3,parker4,parker5,hcurvo1,Barros1,chandrasekharansatz,cohen,neutrino1,santos1}%
. A fundamental question in physics is how quantum systems are affected by
the structure of the spacetime, and if exists some significant effect. In
order to study these gravitational effects, systematic studies \ are being
carried out \cite{santos1,santos2}. For example, in \cite{santos1} where the
effects of very intense magnetic fields in the energy levels, as intense as
the ones expected to be produced in ultra-relativistic heavy-ion collisions,
are investigated, in \cite{santos2} where bosons inside cosmic strings are
considered.

In this paper, a single particle solution of the Dirac equation in an
accelerated reference frame is discussed. The motivation for this work
besides the ones pointed out above is the understanding of the physical
consequences of the Dirac equation in noninertial systems of reference that
undergo translational acceleration.

The paper is organized as follows: section 2 contains a brief review about
the wave equation for spin 1/2 particles in curved spacetimes where the
basic formulation and the equations that will be needed in the next sections
will be shown. In section 3, we will present an equation for spin 1/2
fermions in the Rindler spacetime, i.e., the Dirac equation in an
accelerated reference frame. In section 4, we will see that the energy
spectrum associated with the Dirac equation in Rindler space is discrete and
depends on $a$, the acceleration of the reference frame. Finally, section 5
presents our conclusions. In this work, we use natural units, $\hbar =c=1.$

\section{Fermions in curved spacetimes}

A essential characteristic of the Dirac operator in flat spacetimes is its
invariance under Lorentz transformations, so when these particles are
studied in curved spacetimes, it is necessary to preserve this aspect
(locally). It can be written by using the tetrads $e_{\text{ \ \ \ }\mu
}^{\left( a\right) }$ that may be defined in order to satisfy the expression 
\begin{equation}
g_{\eta \lambda }=e_{\text{ \ \ \ }\eta }^{\left( a\right) }e_{\text{ \ \ \ }%
\lambda }^{\left( b\right) }\eta _{\left( a\right) \left( b\right) },
\label{eq1}
\end{equation}

\noindent here $\eta _{\left( a\right) \left( b\right) }$ is the Minkowski
tensor, and $g_{\eta \lambda }$ the general metric tensor. We use $\left(
a\right) ,\left( b\right) ,\left( c\right) ,...$ to denote the local Lorentz
spacetime and $\alpha ,\beta ,\gamma ,...$ to denote the general spacetime
coordinate \cite{tetrada1,tetrada2,tetrada3}. From eq. $\left( \ref{eq1}%
\right) $, we can see that the tetrads may be used in order to project
vectors from the curved spacetime in the flat spacetime with the equation $%
A_{\mu }=e_{\text{ \ \ \ }\mu }^{\left( a\right) }A_{\left( a\right) }$

\noindent that relates the form of a vector in different spacetimes.

Now we study the behavior of the elements of the Dirac equation under
transformations that preserve the Lorentz symmetry. We note that a spinor
transforms according to $\psi \rightarrow \rho \left( \Lambda \right) \psi $%
, with $\rho \left( \Lambda \right) =1+\frac{1}{2}i\varepsilon ^{\left(
a\right) \left( b\right) }\Sigma _{\left( a\right) \left( b\right) }$, and $%
\Sigma _{\left( a\right) \left( b\right) }$ is the spinoral representation
of the generators of the Lorentz transformation, written in terms of the $%
\gamma ^{\left( c\right) }$ matrices, $\Sigma _{\left( a\right) \left(
b\right) }\equiv \frac{1}{4}i\left[ \gamma _{\left( a\right) },\gamma
_{\left( b\right) }\right] $ \cite{nakahara}. The first task is to construct
a covariant derivative $\nabla _{\left( a\right) }\psi $ that is locally
Lorentz invariant, for this purpose we impose the transformation condition 
\begin{equation}
\nabla _{\left( a\right) }\psi \rightarrow \rho \Lambda _{\left( a\right) }^{%
\text{ \ \ }\left( b\right) }\nabla _{\left( b\right) }\psi .  \label{eq3}
\end{equation}

\noindent The common way to obtain the explicit form of the covariant
derivative operator is by supposing the combination of terms 
\begin{equation}
\nabla _{\left( a\right) }\psi =e_{\left( a\right) }^{\text{ \ \ \ }\mu
}\left( \partial _{\mu }+\Omega _{\mu }\right) \psi ,\text{ }  \label{eq4}
\end{equation}

\noindent we note that the operator $\Omega _{\mu }$ transforms according 
\begin{equation}
\Omega _{\mu }\rightarrow \rho \Omega _{\mu }\rho ^{-1}+\partial _{\mu }\rho
\rho ^{-1}.  \label{eq5}
\end{equation}

The next step is to find the explicit form of the operator $\Omega _{\mu }$.
\noindent If we consider the combination of terms 
\begin{equation}
\Omega _{\mu }\equiv \frac{1}{2}i\Gamma _{\text{ }(a)\mu (b)}\Sigma
^{(a)(b)}=\frac{1}{2}ie_{\text{ \ }\nu }^{\left( a\right) }\nabla _{\mu }e_{%
\text{ \ }}^{\left( b\right) \nu }\Sigma _{\left( a\right) \left( b\right) },
\label{eq6}
\end{equation}

\noindent\ the eq. $\left( \ref{eq5}\right) $ and $\left( \ref{eq3}\right) $
are satisfied, where the term $\Gamma _{\text{ }(a)\mu (b)}$ is given by 
\begin{equation}
\Gamma _{\text{ }(a)\mu (b)}=e_{(a)\nu }\left( \partial _{\mu }e_{(b)}^{%
\text{ \ \ }\nu }+\Gamma _{\text{ \ }\mu \lambda }^{\nu }e_{(b)}^{\text{ \ \ 
}\lambda }\right) ,
\end{equation}

\noindent where $\Gamma _{\text{ \ }\mu \lambda }^{\nu }$ are the
Christoffel symbols. As a result, we obtain the final form of the covariant
derivative operator 
\begin{equation}
\nabla _{\left( c\right) }\psi =e_{\left( c\right) }^{\text{ \ \ \ }\mu
}\left( \partial _{\mu }+\frac{1}{2}ie_{\text{ \ \ \ }\nu }^{\left( a\right)
}\nabla _{\mu }e_{\text{ \ }}^{\left( b\right) \nu }\Sigma _{\left( a\right)
\left( b\right) }\right) \psi .  \label{eq7}
\end{equation}

\noindent By replacing the conventional derivative operator of the Dirac
equation in a flat spacetime by the one obtained in $\left( \ref{eq7}\right) 
$ we obtain the final form of the wave equation for fermions particles in a
curved spacetime 
\begin{equation}
ie_{\left( a\right) }^{\text{ \ \ \ }\mu }\gamma ^{\left( a\right) }\left(
\partial _{\mu }+\Omega _{\mu }\right) \psi -m\psi =0.  \label{eq8}
\end{equation}%
It is common to define the term $\gamma ^{\mu }=e_{\left( a\right) }^{\text{
\ \ \ }\mu }\gamma ^{\left( a\right) }$ as a Dirac matrix in a given curved
spacetime and it is easy to see that it satisfies the Clifford algebra 
\begin{equation}
\gamma ^{\mu }\gamma ^{\nu }+\gamma ^{\nu }\gamma ^{\mu }=g^{\mu \nu }%
\mathbf{1}.  \label{eq10}
\end{equation}

\noindent If the spinor $\psi $ is coupled to the gauge field $A_{\mu }$, we
may introduce this effect by a minimal coupling 
\begin{equation}
ie_{\left( a\right) }^{\text{ \ \ \ }\mu }\gamma ^{\left( a\right) }\left(
\partial _{\mu }+\Omega _{\mu }+ieA_{\mu }\right) \psi -m\psi =0.
\label{eq8b}
\end{equation}

\section*{Wave equation in the Rindler spacetime}

The Rindler metric represents an accelerated reference frame in the
Minkowski spacetime where the line element may be written in the form%
\begin{equation}
ds^{2}=\left( 1+a\xi \right) ^{2}d\tau ^{2}-d\xi ^{2},  \label{eq11}
\end{equation}%
where $\xi $ and $\tau $ are the proper coordinates of the accelerated
frame, and $a$ is the acceleration \cite{qft}. The coordinate of distance $%
\xi $ is constrained by $\xi >-\frac{1}{a}$ while the proper time $\tau $
varies in the interval $-\infty <\tau <\infty $. We shall now\ rewrite the
metric $\left( \ref{eq11}\right) $ in a conformally flat form. The usual way
to obtain the conformal form of eq. $\left( \ref{eq11}\right) $ is by making
the transformation%
\begin{align}
\bar{\tau}& =\tau ,  \notag \\
x& =\frac{1}{a}\ln \left( 1+a\xi \right) .  \label{eq12}
\end{align}%
In this way, we obtain the line element%
\begin{equation}
ds^{2}=e^{2ax}\left( d\bar{\tau}^{2}-dx^{2}\right) .  \label{eq13}
\end{equation}%
Now the conformal coordinates $x$, $\bar{\tau}$ vary in the interval $%
-\infty <\bar{\tau}<\infty ,$ and $-\infty <x<\infty $, that means an
extension of the spacetime metric. From \ eq. $\left( \ref{eq13}\right) $ we
can see that the metric is conformally flat because the conformal factor $%
e^{2ax}$ is multiplied by the Minkowski line element. In addition, the
spacetime metric $\left( \ref{eq13}\right) $ represents a flat spacetime
since it is related to the Minkowski line element by a coordinate
transformation.

Thus, the next step is to choose the tetrad basis for the line element $%
\left( \ref{eq13}\right) $. The diagonal form of $\left( \ref{eq13}\right) $
suggests the following tetrad basis 
\begin{align}
e_{\nu }^{\text{ \ }\left( a\right) }& =\left( 
\begin{array}{c}
e^{\frac{\sigma \left( x\right) }{2}} \\ 
0%
\end{array}%
\begin{array}{c}
0 \\ 
e^{\frac{\sigma \left( x\right) }{2}}%
\end{array}%
\right) ,  \label{eq15} \\
e_{\text{ \ }\left( a\right) }^{\nu }& =\left( 
\begin{array}{c}
e^{-\frac{\sigma \left( x\right) }{2}} \\ 
0%
\end{array}%
\begin{array}{c}
0 \\ 
e^{-\frac{\sigma \left( x\right) }{2}}%
\end{array}%
\right) ,  \label{eq16}
\end{align}%
where $\sigma \left( x\right) \equiv 2ax$. It is easy to verify that it
satisfies the equation%
\begin{equation*}
g_{\mu \nu }=e_{\text{ \ \ \ }\mu }^{\left( a\right) }e_{\text{ \ \ \ }\nu
}^{\left( b\right) }\eta _{\left( a\right) \left( b\right) }.
\end{equation*}%
Observing that the only non-zero term $\Omega _{0}$ in the wave equation,
relative to the tetrad (\ref{eq15}) is given by%
\begin{equation}
\Omega _{0}=\frac{i\gamma ^{0}\cdot \gamma ^{1}}{4}\frac{d\sigma }{dx},
\label{eq17}
\end{equation}%
equation (\ref{eq8}) becomes%
\begin{equation}
\left[ i\gamma ^{\left( 0\right) }\frac{\partial }{\partial \bar{\tau}}%
+i\gamma ^{\left( 1\right) }\frac{\partial }{\partial x}+\frac{i\gamma
^{\left( 1\right) }}{4}\frac{d\sigma \left( x\right) }{dx}-me^{\frac{1}{2}%
\sigma \left( x\right) }\right] \psi \left( \bar{\tau},x\right) =0,
\label{eq18}
\end{equation}%
where the usual representation for the gamma matrices in $1+1$ dimensions is
considered%
\begin{equation}
\gamma ^{\left( 0\right) }=\left( 
\begin{array}{c}
0 \\ 
1%
\end{array}%
\begin{array}{c}
1 \\ 
0%
\end{array}%
\right) ,\text{ }\gamma ^{\left( 1\right) }=\left( 
\begin{array}{c}
i \\ 
0%
\end{array}%
\begin{array}{c}
0 \\ 
-i%
\end{array}%
\right) .  \label{eq19}
\end{equation}%
In eq.$\left( \ref{eq18}\right) ,$ we will suppose a solution of the form%
\begin{equation}
\psi \left( t,x\right) =e^{-i\varepsilon \bar{\tau}}\left[ 
\begin{array}{c}
\bar{g}\left( x\right) \\ 
\bar{f}\left( x\right)%
\end{array}%
\right] ,  \label{eq22}
\end{equation}%
where $\varepsilon $ is the energy of particle. So, equation (\ref{eq18})
may be written in an explicit form%
\begin{align}
\left[ \frac{d}{dx}+\frac{1}{4}\frac{d\sigma \left( x\right) }{dx}+me^{\frac{%
1}{2}\sigma \left( x\right) }\right] \bar{g}\left( x\right) & =\varepsilon 
\bar{f}\left( x\right) ,  \label{eq23} \\
\left[ -\frac{d}{dx}-\frac{1}{4}\frac{d\sigma \left( x\right) }{dx}+me^{%
\frac{1}{2}\sigma \left( x\right) }\right] \bar{f}\left( x\right) &
=\varepsilon \bar{g}\left( x\right) .  \label{eq24}
\end{align}%
Making a unitary transformation \cite{diraco14} in the system of equations%
\begin{equation}
U\left( x\right) =\left[ 
\begin{array}{c}
e^{\frac{1}{4}\sigma \left( x\right) } \\ 
0%
\end{array}%
\begin{array}{c}
0 \\ 
e^{\frac{1}{4}\sigma \left( x\right) }%
\end{array}%
\right] ,  \label{eq25}
\end{equation}%
we obtain a simplified form%
\begin{align}
\left[ \frac{d}{dx}+z\left( x\right) \right] g\left( x\right) & =\varepsilon
f\left( x\right) ,  \label{eq26} \\
\left[ -\frac{d}{dx}+z\left( x\right) \right] f\left( x\right) &
=\varepsilon g\left( x\right) ,  \label{eq27}
\end{align}%
where $z\left( x\right) =me^{\frac{1}{2}\sigma \left( x\right) }$ and%
\begin{equation}
\bar{\psi}=U\psi =U\left[ 
\begin{array}{c}
\bar{g}\left( x\right) \\ 
\bar{f}\left( x\right)%
\end{array}%
\right] =\left[ 
\begin{array}{c}
g\left( x\right) \\ 
f\left( x\right)%
\end{array}%
\right] .  \label{eq28}
\end{equation}%
In order to investigate the solutions of the system of equations, we will
consider that the set of equations $\left( \ref{eq26}\right) $ and $\left( %
\ref{eq27}\right) $ may be decoupled. The function $f$ \ can be isolated in
eq. $\left( \ref{eq26}\right) $ and then, using $\left( \ref{eq26}\right) $
to eliminate the terms containing the $g$. The result is%
\begin{equation}
\frac{d^{2}f}{dx^{2}}-\frac{dz}{dx}f-z^{2}f+\varepsilon ^{2}f=0,
\label{eq29}
\end{equation}%
in a similar way, we derive the equation for $g$%
\begin{equation}
\frac{d^{2}g}{dx^{2}}+\frac{dz}{dx}g-z^{2}g+\varepsilon ^{2}g=0,
\label{eq30}
\end{equation}%
that may be resumed in the form%
\begin{equation}
\frac{d^{2}F}{dx^{2}}-s\frac{dz}{dx}F-z^{2}F+\varepsilon ^{2}F=0,
\label{eq31}
\end{equation}%
where $F=f$ when $s=1$ and $F=g$ when $s=-1$. For the case of the Rindler
metric, we have $\sigma =2ax$ so that%
\begin{equation}
g_{\mu \eta }=\exp \left( 2ax\right) \left( 
\begin{array}{cc}
1 & 0 \\ 
0 & -1%
\end{array}%
\right) .  \label{eq32}
\end{equation}%
If we consider small values of the acceleration $a$, we may solve eq. $%
\left( \ref{eq31}\right) $ neglecting terms in higher orders, and then $%
z\left( x\right) $ can be expanded as%
\begin{equation}
z\left( x\right) =m\left( 1+\frac{ax}{2}+...\right) .  \label{eq33}
\end{equation}%
So, substituting this equation into Eq. (\ref{eq31}) we obtain the following
expression%
\begin{equation}
\frac{d^{2}F}{dx^{2}}-s\frac{ma}{2}F-\frac{a^{2}m^{2}}{4}\left( \frac{2}{a}%
+x\right) ^{2}F+\varepsilon ^{2}F=0.  \label{eq34}
\end{equation}%
In order to investigate the solutions of the Eq. (\ref{eq34}), we will
consider the transformation 
\begin{equation}
y=\sqrt{\frac{am}{2}}\left( x+\frac{2}{a}\right) ,  \label{eq35}
\end{equation}%
and as a result, the equation will take the form%
\begin{equation}
\frac{d^{2}F}{dy^{2}}+\left( \eta -V_{ef}\right) =0,  \label{eq36}
\end{equation}%
here we have defined $\eta =\frac{2\varepsilon ^{2}}{am}$ end $%
V_{ef}=y^{2}+s.$ In the next section we will obtain two classes of solutions
of the Dirac equation in the Rindler spacetime. Indeed, eq. $\left( \ref%
{eq36}\right) $ may be mapped into a Sturm--Liouville problem of a Schr\"{o}%
dinger-like equation.

\section{Bound-state solutions}

As we may observe, equation (\ref{eq36}) is similar to the Schr\"{o}dinger
equation, and the term $V_{ef}=y^{2}+s$ may be identified as an effective
potential and as we can see, the system has the form of a harmonic
oscillator. In fact, the solution of the equation (\ref{eq36}) may be mapped
into a 2-dimensional harmonic oscillator-like, and then the

\begin{figure}[tbp]
\includegraphics[scale=0.6]{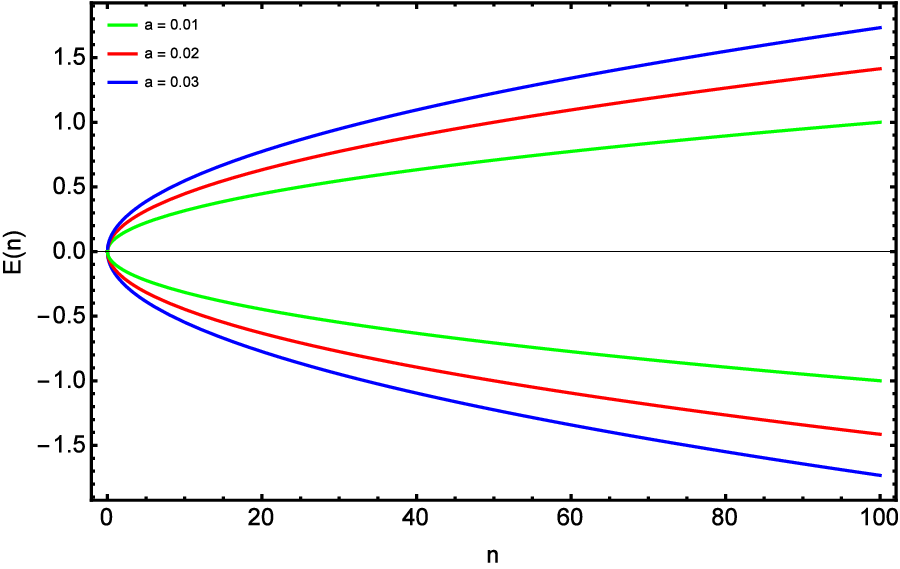}\newline
\caption{The plots of the energy spectrum for $a=0.01,a=0.02$ and $a=0.03.$
The energy is symmetrical about $\protect\varepsilon =0.$}
\label{fig1}
\end{figure}
solutions are given in terms of the Hermite polynomials

\begin{equation}
\bar{\psi}=\left( 
\begin{array}{c}
B_{1}^{\prime }\exp \left( -y^{2}\right) H_{n+1}\left( y\right) \\ 
B_{\substack{ 2  \\ }}^{\prime }\exp \left( -y^{2}\right) H_{n}\left(
y\right)%
\end{array}%
\right) ,  \label{eq37}
\end{equation}%
where $B_{1}^{\prime }$ and $B_{2}^{\prime }$ are normalization constants.
The Hermite polynomials satisfy the recurrence relation%
\begin{equation}
a_{j+2}=\frac{2j+1-\eta }{\left( j+2\right) \left( j+1\right) }a_{j},
\label{eq38}
\end{equation}%
as the series must be finite in order for our solution to have physical
meaning, we suppose that there exists some $n$ such that when $j=n$, the
numerator is $2j+1-\eta =0$. In this way, we obtain the energy spectrum%
\begin{equation}
\varepsilon =\pm \sqrt{am\left( n+\frac{1+s}{2}\right) .}  \label{eq39}
\end{equation}

We can see that the energy spectrum associated with the Dirac equation in
the Rindler space is discrete and depends on $a$, the acceleration of the
reference frame. This is a interesting feature of the system because the
noninertial effect mimics an external potential in the Dirac equation. From
Fig. 1, we can see that the discrete set of energies are symmetrical about $%
\varepsilon =0$, that means the particle and antiparticle have the same
energy. As it may be seen in Fig. 2, the solution $g\left( x\right) $
decreases with the coordinate $x$ and becomes negligible as $x\rightarrow
\pm \infty $.

\begin{figure}[H]
\includegraphics[scale=0.75]{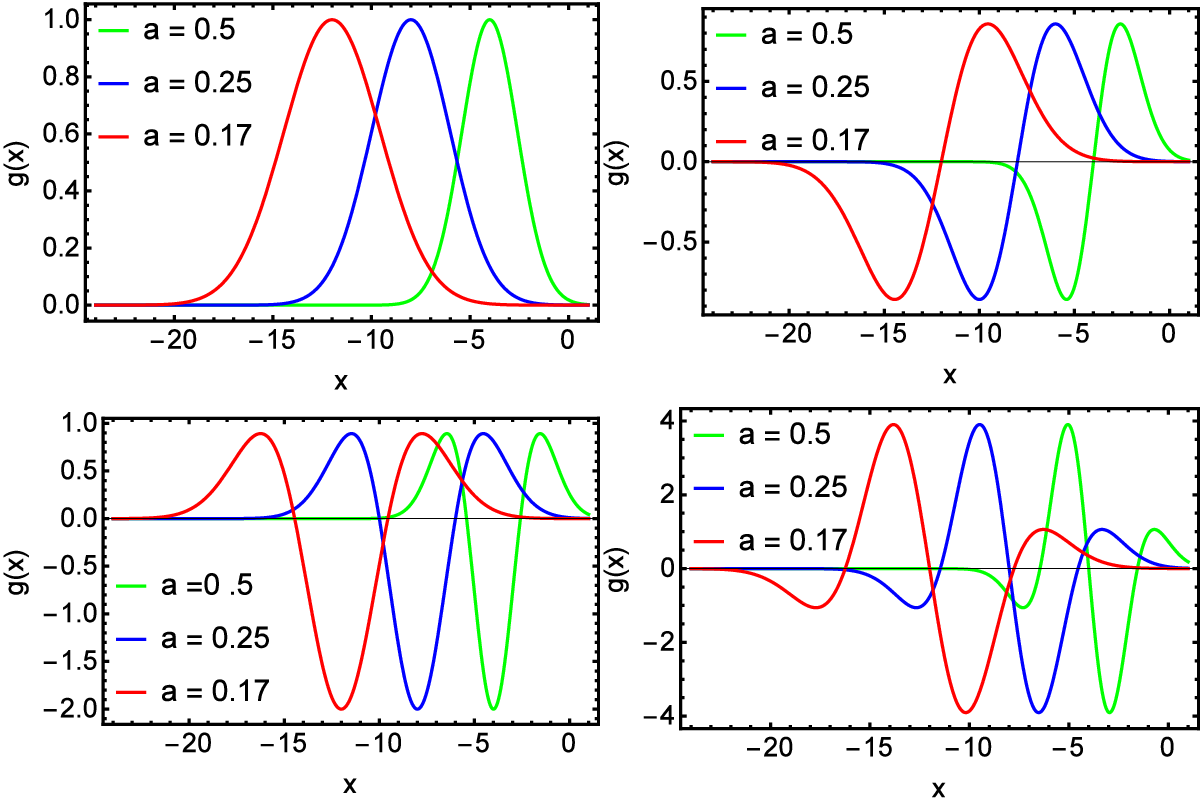}\newline
\caption{The lower spinor component $g\left( x\right) $ as the function of $%
x $ for three fixed values of $a,$ Up: left and right the figures are plots
for $n=0$ and $n=1$. Down: left and right the figures are plots for $n=2$
and $n=3.$}
\label{fig2}
\end{figure}
\section{Conclusions}

In this work a brief review about the wave equation for spin 1/2 particles
in curved spaces is done. We have determined a single particle solution of
the Dirac equation in an accelerated reference frame, and as result, a
compact expression for the energy spectrum associated with the Dirac
equation in an accelerated reference frame has been obtained.

It was shown that the noninertial effect mimics an external potential in the
Dirac equation and, moreover, allows the formation of bound states. We also
have shown that the energy spectrum associated with fermions in this kind
spacetime is discrete. The solution is obtained by adopting the limit $a\ll
1 $, that means a not so fast acceleration.

With these results it is possible to have an idea about the general aspects
of the behavior of spin 1/2 particles in the Rindler space. Potential
applications of our work include physical systems with conformally flat
metrics, i.e, invariant under conformal transformations. For instance, in
applications of the AdS/CFT correspondence \cite{adscft1} to the study of
strongly coupled QCD, we look for solutions of the wave equations where the
conformal symmetry plays a pivotal role \cite{adscft2}.

It is interesting to observe that the results obtained above, in addition to
previous ones \cite{santos1,santos2}, for example, show many important
aspects of quantum systems studied in spacetimes with different structures.
However, these results rather than being considered as final, may be
considered as a motivation for future works in order to obtain a deeper
understanding of this fundamental theme of physics.

\section{Acknowledgments}

This work was supported in part by means of funds provided by CNPq
\begin{equation*}
\end{equation*}%
\bigskip

\bigskip 
\bibliographystyle{aipnum4-1}
\bibliography{C:/referencias/referencias_unificadas}

\begin{thebibliography}{28}%
\makeatletter
\providecommand \@ifxundefined [1]{%
 \@ifx{#1\undefined}
}%
\providecommand \@ifnum [1]{%
 \ifnum #1\expandafter \@firstoftwo
 \else \expandafter \@secondoftwo
 \fi
}%
\providecommand \@ifx [1]{%
 \ifx #1\expandafter \@firstoftwo
 \else \expandafter \@secondoftwo
 \fi
}%
\providecommand \natexlab [1]{#1}%
\providecommand \enquote  [1]{``#1''}%
\providecommand \bibnamefont  [1]{#1}%
\providecommand \bibfnamefont [1]{#1}%
\providecommand \citenamefont [1]{#1}%
\providecommand \href@noop [0]{\@secondoftwo}%
\providecommand \href [0]{\begingroup \@sanitize@url \@href}%
\providecommand \@href[1]{\@@startlink{#1}\@@href}%
\providecommand \@@href[1]{\endgroup#1\@@endlink}%
\providecommand \@sanitize@url [0]{\catcode `\\12\catcode `\$12\catcode
  `\&12\catcode `\#12\catcode `\^12\catcode `\_12\catcode `\%12\relax}%
\providecommand \@@startlink[1]{}%
\providecommand \@@endlink[0]{}%
\providecommand \url  [0]{\begingroup\@sanitize@url \@url }%
\providecommand \@url [1]{\endgroup\@href {#1}{\urlprefix }}%
\providecommand \urlprefix  [0]{URL }%
\providecommand \Eprint [0]{\href }%
\providecommand \doibase [0]{http://dx.doi.org/}%
\providecommand \selectlanguage [0]{\@gobble}%
\providecommand \bibinfo  [0]{\@secondoftwo}%
\providecommand \bibfield  [0]{\@secondoftwo}%
\providecommand \translation [1]{[#1]}%
\providecommand \BibitemOpen [0]{}%
\providecommand \bibitemStop [0]{}%
\providecommand \bibitemNoStop [0]{.\EOS\space}%
\providecommand \EOS [0]{\spacefactor3000\relax}%
\providecommand \BibitemShut  [1]{\csname bibitem#1\endcsname}%
\let\auto@bib@innerbib\@empty
\bibitem [{\citenamefont {Castro}(2016)}]{Castro1}%
  \BibitemOpen
  \bibfield  {author} {\bibinfo {author} {\bibfnamefont {L.~B.}\ \bibnamefont
  {Castro}},\ }\href {\doibase 10.1140/epjc/s10052-016-3904-4} {\bibfield
  {journal} {\bibinfo  {journal} {Eur. Phys. J. C}\ }\textbf {\bibinfo {volume}
  {76}},\ \bibinfo {pages} {1} (\bibinfo {year} {2016})}\BibitemShut {NoStop}%
\bibitem [{\citenamefont {Castro}(2015)}]{Castro2}%
  \BibitemOpen
  \bibfield  {author} {\bibinfo {author} {\bibfnamefont {L.~B.}\ \bibnamefont
  {Castro}},\ }\href {\doibase 10.1140/epjc/s10052-015-3507-5} {\bibfield
  {journal} {\bibinfo  {journal} {Eur. Phys. J. C}\ }\textbf {\bibinfo {volume}
  {75}},\ \bibinfo {pages} {1} (\bibinfo {year} {2015})}\BibitemShut {NoStop}%
\bibitem [{\citenamefont {Shen}, \citenamefont {He},\ and\ \citenamefont
  {Zhuang}(2005)}]{inercial9}%
  \BibitemOpen
  \bibfield  {author} {\bibinfo {author} {\bibfnamefont {J.~Q.}\ \bibnamefont
  {Shen}}, \bibinfo {author} {\bibfnamefont {S.}~\bibnamefont {He}}, \ and\
  \bibinfo {author} {\bibfnamefont {F.}~\bibnamefont {Zhuang}},\ }\href
  {\doibase 10.1140/epjd/e2005-00027-7} {\bibfield  {journal} {\bibinfo
  {journal} {Eur. Phys. J. D}\ }\textbf {\bibinfo {volume} {33}},\ \bibinfo
  {pages} {35} (\bibinfo {year} {2005})}\BibitemShut {NoStop}%
\bibitem [{\citenamefont {Bakke}(2013)}]{inercial10}%
  \BibitemOpen
  \bibfield  {author} {\bibinfo {author} {\bibfnamefont {K.}~\bibnamefont
  {Bakke}},\ }\href {\doibase 10.1007/s10714-013-1561-6} {\bibfield  {journal}
  {\bibinfo  {journal} {Gen.Rel.Grav.}\ }\textbf {\bibinfo {volume} {45}},\
  \bibinfo {pages} {1847} (\bibinfo {year} {2013})},\ \Eprint
  {http://arxiv.org/abs/1307.2847} {arXiv:1307.2847 [quant-ph]} \BibitemShut
  {NoStop}%
\bibitem [{\citenamefont {Werner}, \citenamefont {Staudenmann},\ and\
  \citenamefont {Colella}(1979)}]{inercial7}%
  \BibitemOpen
  \bibfield  {author} {\bibinfo {author} {\bibfnamefont {S.}~\bibnamefont
  {Werner}}, \bibinfo {author} {\bibfnamefont {J.-L.}\ \bibnamefont
  {Staudenmann}}, \ and\ \bibinfo {author} {\bibfnamefont {R.}~\bibnamefont
  {Colella}},\ }\href@noop {} {\bibfield  {journal} {\bibinfo  {journal} {Phys.
  Rev. Lett.}\ }\textbf {\bibinfo {volume} {42}},\ \bibinfo {pages} {1103}
  (\bibinfo {year} {1979})}\BibitemShut {NoStop}%
\bibitem [{\citenamefont {Fulling}(1973)}]{uru1}%
  \BibitemOpen
  \bibfield  {author} {\bibinfo {author} {\bibfnamefont {S.}~\bibnamefont
  {Fulling}},\ }\href {\doibase 10.1103/PhysRevD.7.2850} {\bibfield  {journal}
  {\bibinfo  {journal} {Phys. Rev. D}\ }\textbf {\bibinfo {volume} {7}},\
  \bibinfo {pages} {2850} (\bibinfo {year} {1973})}\BibitemShut {NoStop}%
\bibitem [{\citenamefont {Davies}(1975)}]{uru2}%
  \BibitemOpen
  \bibfield  {author} {\bibinfo {author} {\bibfnamefont {P.~C.~W.}\
  \bibnamefont {Davies}},\ }\href {http://stacks.iop.org/0305-4470/8/i=4/a=022}
  {\bibfield  {journal} {\bibinfo  {journal} {J. Phys. A: Math. Gen.}\ }\textbf
  {\bibinfo {volume} {8}},\ \bibinfo {pages} {609} (\bibinfo {year}
  {1975})}\BibitemShut {NoStop}%
\bibitem [{\citenamefont {Unruh}(1976)}]{uru3}%
  \BibitemOpen
  \bibfield  {author} {\bibinfo {author} {\bibfnamefont {W.}~\bibnamefont
  {Unruh}},\ }\href {\doibase 10.1103/PhysRevD.14.870} {\bibfield  {journal}
  {\bibinfo  {journal} {Phys. Rev. D}\ }\textbf {\bibinfo {volume} {14}},\
  \bibinfo {pages} {870} (\bibinfo {year} {1976})}\BibitemShut {NoStop}%
\bibitem [{\citenamefont {Parker}(1980{\natexlab{a}})}]{parker1}%
  \BibitemOpen
  \bibfield  {author} {\bibinfo {author} {\bibfnamefont {L.}~\bibnamefont
  {Parker}},\ }\href@noop {} {\bibfield  {journal} {\bibinfo  {journal} {Phys.
  Rev. Lett.}\ }\textbf {\bibinfo {volume} {44}},\ \bibinfo {pages} {1559}
  (\bibinfo {year} {1980}{\natexlab{a}})}\BibitemShut {NoStop}%
\bibitem [{\citenamefont {Parker}(1980{\natexlab{b}})}]{parker2}%
  \BibitemOpen
  \bibfield  {author} {\bibinfo {author} {\bibfnamefont {L.}~\bibnamefont
  {Parker}},\ }\href@noop {} {\bibfield  {journal} {\bibinfo  {journal} {Phys.
  Rev. D}\ }\textbf {\bibinfo {volume} {22}},\ \bibinfo {pages} {1922}
  (\bibinfo {year} {1980}{\natexlab{b}})}\BibitemShut {NoStop}%
\bibitem [{\citenamefont {Parker}(1981{\natexlab{a}})}]{parker3}%
  \BibitemOpen
  \bibfield  {author} {\bibinfo {author} {\bibfnamefont {L.}~\bibnamefont
  {Parker}},\ }\href@noop {} {\bibfield  {journal} {\bibinfo  {journal} {Phys.
  Rev. D}\ }\textbf {\bibinfo {volume} {24}},\ \bibinfo {pages} {535} (\bibinfo
  {year} {1981}{\natexlab{a}})}\BibitemShut {NoStop}%
\bibitem [{\citenamefont {Parker}(1981{\natexlab{b}})}]{parker4}%
  \BibitemOpen
  \bibfield  {author} {\bibinfo {author} {\bibfnamefont {L.}~\bibnamefont
  {Parker}},\ }\href@noop {} {\bibfield  {journal} {\bibinfo  {journal}
  {Gen.Rel.Grav.}\ }\textbf {\bibinfo {volume} {13}},\ \bibinfo {pages} {307}
  (\bibinfo {year} {1981}{\natexlab{b}})}\BibitemShut {NoStop}%
\bibitem [{\citenamefont {Parker}\ and\ \citenamefont
  {Pimentel}(1982)}]{parker5}%
  \BibitemOpen
  \bibfield  {author} {\bibinfo {author} {\bibfnamefont {L.}~\bibnamefont
  {Parker}}\ and\ \bibinfo {author} {\bibfnamefont {L.}~\bibnamefont
  {Pimentel}},\ }\href@noop {} {\bibfield  {journal} {\bibinfo  {journal}
  {Phys. Rev. D}\ }\textbf {\bibinfo {volume} {25}},\ \bibinfo {pages} {3180}
  (\bibinfo {year} {1982})}\BibitemShut {NoStop}%
\bibitem [{\citenamefont {Marques}\ and\ \citenamefont
  {Bezerra}(2002)}]{hcurvo1}%
  \BibitemOpen
  \bibfield  {author} {\bibinfo {author} {\bibfnamefont {G.}~\bibnamefont
  {Marques}}\ and\ \bibinfo {author} {\bibfnamefont {V.}~\bibnamefont
  {Bezerra}},\ }\href
  {http://www.scopus.com/inward/record.url?eid=2-s2.0-0242267817&partnerID=40&md5=f9f8c34334692b6247486472e2bd802d}
  {\bibfield  {journal} {\bibinfo  {journal} {Phys. Rev. D}\ }\textbf {\bibinfo
  {volume} {66}},\ \bibinfo {pages} {105011} (\bibinfo {year}
  {2002})}\BibitemShut {NoStop}%
\bibitem [{\citenamefont {Barros}(2005)}]{Barros1}%
  \BibitemOpen
  \bibfield  {author} {\bibinfo {author} {\bibfnamefont {C.~C.}\ \bibnamefont
  {Barros}},\ }\href@noop {} {\bibfield  {journal} {\bibinfo  {journal} {Eur.
  Phys. J. C}\ }\textbf {\bibinfo {volume} {42}},\ \bibinfo {pages} {119}
  (\bibinfo {year} {2005})}\BibitemShut {NoStop}%
\bibitem [{\citenamefont {Chandrasekhar}(1976)}]{chandrasekharansatz}%
  \BibitemOpen
  \bibfield  {author} {\bibinfo {author} {\bibfnamefont {S.}~\bibnamefont
  {Chandrasekhar}},\ }\href@noop {} {\bibfield  {journal} {\bibinfo  {journal}
  {Proceedings of the Royal Society of London. Series A, Mathematical and
  Physical Sciences}\ }\textbf {\bibinfo {volume} {349}},\ \bibinfo {pages}
  {571} (\bibinfo {year} {1976})}\BibitemShut {NoStop}%
\bibitem [{\citenamefont {Cohen}\ and\ \citenamefont {Powers}(1982)}]{cohen}%
  \BibitemOpen
  \bibfield  {author} {\bibinfo {author} {\bibfnamefont {J.~M.}\ \bibnamefont
  {Cohen}}\ and\ \bibinfo {author} {\bibfnamefont {R.~T.}\ \bibnamefont
  {Powers}},\ }\href@noop {} {\bibfield  {journal} {\bibinfo  {journal}
  {Communications in Mathematical Physics}\ }\textbf {\bibinfo {volume} {86}},\
  \bibinfo {pages} {69} (\bibinfo {year} {1982})}\BibitemShut {NoStop}%
\bibitem [{\citenamefont {Brill}\ and\ \citenamefont
  {Wheeler}(1957)}]{neutrino1}%
  \BibitemOpen
  \bibfield  {author} {\bibinfo {author} {\bibfnamefont {D.~R.}\ \bibnamefont
  {Brill}}\ and\ \bibinfo {author} {\bibfnamefont {J.~A.}\ \bibnamefont
  {Wheeler}},\ }\href {\doibase 10.1103/RevModPhys.29.465} {\bibfield
  {journal} {\bibinfo  {journal} {Rev.Mod.Phys.}\ }\textbf {\bibinfo {volume}
  {29}},\ \bibinfo {pages} {465} (\bibinfo {year} {1957})}\BibitemShut
  {NoStop}%
\bibitem [{\citenamefont {Santos}\ and\ \citenamefont
  {Barros}(2016)}]{santos1}%
  \BibitemOpen
  \bibfield  {author} {\bibinfo {author} {\bibfnamefont {L.~C.~N.}\
  \bibnamefont {Santos}}\ and\ \bibinfo {author} {\bibfnamefont {C.~C.}\
  \bibnamefont {Barros}},\ }\href@noop {} {\bibfield  {journal} {\bibinfo
  {journal} {Eur. Phys. J. C}\ }\textbf {\bibinfo {volume} {76}},\ \bibinfo
  {pages} {560} (\bibinfo {year} {2016})}\BibitemShut {NoStop}%
\bibitem [{\citenamefont {Santos}\ and\ \citenamefont
  {Barros}(2017)}]{santos2}%
  \BibitemOpen
  \bibfield  {author} {\bibinfo {author} {\bibfnamefont {L.~C.~N.}\
  \bibnamefont {Santos}}\ and\ \bibinfo {author} {\bibfnamefont {C.~C.}\
  \bibnamefont {Barros}},\ }\href {\doibase 10.1140/epjc/s10052-017-4732-x}
  {\bibfield  {journal} {\bibinfo  {journal} {Eur. Phys. J. C}\ }\textbf
  {\bibinfo {volume} {77}},\ \bibinfo {pages} {186} (\bibinfo {year}
  {2017})}\BibitemShut {NoStop}%
\bibitem [{\citenamefont {Obukhov}, \citenamefont {Silenko},\ and\
  \citenamefont {Teryaev}(2013)}]{tetrada1}%
  \BibitemOpen
  \bibfield  {author} {\bibinfo {author} {\bibfnamefont {Y.~N.}\ \bibnamefont
  {Obukhov}}, \bibinfo {author} {\bibfnamefont {A.~J.}\ \bibnamefont
  {Silenko}}, \ and\ \bibinfo {author} {\bibfnamefont {O.~V.}\ \bibnamefont
  {Teryaev}},\ }\href {\doibase 10.1103/PhysRevD.88.084014} {\bibfield
  {journal} {\bibinfo  {journal} {Phys. Rev. D}\ }\textbf {\bibinfo {volume}
  {88}},\ \bibinfo {pages} {084014} (\bibinfo {year} {2013})}\BibitemShut
  {NoStop}%
\bibitem [{\citenamefont {Obukhov}, \citenamefont {Silenko},\ and\
  \citenamefont {Teryaev}(2009)}]{tetrada2}%
  \BibitemOpen
  \bibfield  {author} {\bibinfo {author} {\bibfnamefont {Y.~N.}\ \bibnamefont
  {Obukhov}}, \bibinfo {author} {\bibfnamefont {A.~J.}\ \bibnamefont
  {Silenko}}, \ and\ \bibinfo {author} {\bibfnamefont {O.~V.}\ \bibnamefont
  {Teryaev}},\ }\href {\doibase 10.1103/PhysRevD.80.064044} {\bibfield
  {journal} {\bibinfo  {journal} {Phys. Rev. D}\ }\textbf {\bibinfo {volume}
  {80}},\ \bibinfo {pages} {064044} (\bibinfo {year} {2009})}\BibitemShut
  {NoStop}%
\bibitem [{\citenamefont {Schwinger}(1963)}]{tetrada3}%
  \BibitemOpen
  \bibfield  {author} {\bibinfo {author} {\bibfnamefont {J.}~\bibnamefont
  {Schwinger}},\ }\href {\doibase 10.1103/PhysRev.130.800} {\bibfield
  {journal} {\bibinfo  {journal} {Phys. Rev.}\ }\textbf {\bibinfo {volume}
  {130}},\ \bibinfo {pages} {800} (\bibinfo {year} {1963})}\BibitemShut
  {NoStop}%
\bibitem [{\citenamefont {Nakahara}(2003)}]{nakahara}%
  \BibitemOpen
  \bibfield  {author} {\bibinfo {author} {\bibfnamefont {M.}~\bibnamefont
  {Nakahara}},\ }\href@noop {} {\emph {\bibinfo {title} {Geometry, Topology and
  Physics, Second Edition}}},\ Graduate student series in physics\ (\bibinfo
  {publisher} {Taylor \& Francis},\ \bibinfo {year} {2003})\BibitemShut
  {NoStop}%
\bibitem [{\citenamefont {Mukhanov}\ and\ \citenamefont
  {Winitzki}(2007)}]{qft}%
  \BibitemOpen
  \bibfield  {author} {\bibinfo {author} {\bibfnamefont {V.}~\bibnamefont
  {Mukhanov}}\ and\ \bibinfo {author} {\bibfnamefont {S.}~\bibnamefont
  {Winitzki}},\ }\href@noop {} {\emph {\bibinfo {title} {Introduction to
  quantum effects in gravity}}}\ (\bibinfo  {publisher} {Cambridge University
  Press},\ \bibinfo {year} {2007})\BibitemShut {NoStop}%
\bibitem [{\citenamefont {Moayedi}\ and\ \citenamefont
  {Darabi}(2004)}]{diraco14}%
  \BibitemOpen
  \bibfield  {author} {\bibinfo {author} {\bibfnamefont {S.}~\bibnamefont
  {Moayedi}}\ and\ \bibinfo {author} {\bibfnamefont {F.}~\bibnamefont
  {Darabi}},\ }\href@noop {} {\bibfield  {journal} {\bibinfo  {journal}
  {Physics Letters A}\ }\textbf {\bibinfo {volume} {322}},\ \bibinfo {pages}
  {173} (\bibinfo {year} {2004})}\BibitemShut {NoStop}%
\bibitem [{\citenamefont {Maldacena}\ \emph {et~al.}(1999)\citenamefont
  {Maldacena}, \citenamefont {Gamboa~Saravi}, \citenamefont {Falomir},\ and\
  \citenamefont {Schaposnik}}]{adscft1}%
  \BibitemOpen
  \bibfield  {author} {\bibinfo {author} {\bibfnamefont {J.}~\bibnamefont
  {Maldacena}}, \bibinfo {author} {\bibfnamefont {R.~E.}\ \bibnamefont
  {Gamboa~Saravi}}, \bibinfo {author} {\bibfnamefont {H.}~\bibnamefont
  {Falomir}}, \ and\ \bibinfo {author} {\bibfnamefont {F.~A.}\ \bibnamefont
  {Schaposnik}},\ }in\ \href@noop {} {\emph {\bibinfo {booktitle} {AIP
  Conference Proceedings CONF-981170}}},\ Vol.\ \bibinfo {volume} {484}\
  (\bibinfo {organization} {AIP},\ \bibinfo {year} {1999})\ pp.\ \bibinfo
  {pages} {51--63}\BibitemShut {NoStop}%
\bibitem [{\citenamefont {Karch}\ \emph {et~al.}(2006)\citenamefont {Karch},
  \citenamefont {Katz}, \citenamefont {Son},\ and\ \citenamefont
  {Stephanov}}]{adscft2}%
  \BibitemOpen
  \bibfield  {author} {\bibinfo {author} {\bibfnamefont {A.}~\bibnamefont
  {Karch}}, \bibinfo {author} {\bibfnamefont {E.}~\bibnamefont {Katz}},
  \bibinfo {author} {\bibfnamefont {D.~T.}\ \bibnamefont {Son}}, \ and\
  \bibinfo {author} {\bibfnamefont {M.~A.}\ \bibnamefont {Stephanov}},\
  }\href@noop {} {\bibfield  {journal} {\bibinfo  {journal} {Phy. Rev. D}\
  }\textbf {\bibinfo {volume} {74}},\ \bibinfo {pages} {015005} (\bibinfo
  {year} {2006})}\BibitemShut {NoStop}%
\end{thebibliography}%

\end{document}